\documentstyle[aps,prl,twocolumn,epsfig]{revtex}

\begin{document}

\def\BE{\begin{equation}}
\def\EE{\end{equation}}
\def\BA{\begin{array}}
\def\EA{\end{array}}
\def\r{\vec{\rho}}

\title{Quantum holographic teleportation of light fields}
\author{I.~V.~Sokolov$^{1}$, M.~I.~Kolobov$^{2}$, A.~Gatti$^{3}$, and L.~A.~
Lugiato$^{3}$}
\address{$^1$ Physics Institute, St.~Petersburg University, 198904
Petrodvorets, St.~Petersburg, Russia}
\address{$^2$ Fachbereich Physik, Universit\"at-GH Essen,
D-45117 Essen, Germany}
\address{$^3$ Dipartimento di Fisica, INFM, Via Celoria 16, 20133
Milano, Italy}
\date{\today}
\maketitle

\begin{abstract}
We describe a continuous variable teleportation scheme that allows to
teleport with high fidelity the quantum state of broadband multimode
electromagnetic field. We call this scheme ``quantum holographic 
teleportation'' because it allows for reconstruction of an optical wavefront 
preserving its quantum correlations in space-time.
\end{abstract}

Quantum teleportation allows for transportation of an arbitrary quantum state 
of
a field from one place to another using classical information exchange.
Initially proposed for discrete variables\cite{Bennett93}, later on
teleportation was extended to continuous-variable
schemes\cite{Vaidman94,Braunstein98a}. Experimental demonstration of quantum
teleportation for discrete variables was realized in
\cite{Bouwmeester97,Boschi98}
for single-photon polarization states, and continuous-variable teleportation
in\cite{Furusawa98} for coherent state of electromagnetic field. Next
challenging task is teleportation of truly nonclassical states like entangled
states or so-called ``entanglement swapping''. The concept of entanglement
swapping was initially introduced for single-photon polarization
states\cite{Zukowski93} and has already been realized experimentally for single
photons\cite{Pan98}. There are several proposals of entanglement swapping for
continuous-variable teleportation schemes\cite{Polkinghorne99,Tan99,vanLoock99}.
Apart from obvious interest to teleportation due to its fundamental quantum
nature nonexistent in classical physics, there is a practical interest to this
phenomenon stirred by potential applications in quantum error
correction\cite{Braunstein98b,Lloyd98,Braunstein98c}, quantum dence
coding\cite{Braunstein00}, and quantum cryptography\cite{Ralph99}.

To date, most theoretical schemes consider quantum teleportation of just a
single-mode state of the field. Such an assumption greatly simplifies analysis
of the teleportation protocol and calculation of parameters describing the
performance of the scheme. However, in reality one has to deal with optical
signals distributed in space-time which have characteristic spatio-temporal
scales such as coherence time and coherence area, for example. To understand the
role of these parameters in teleportation process we have to abandon the
single-mode approximation and generalize quantum teleportation for multimode
states of electromagnetic field.

While broadband teleportation of time-dependent signals has been already
discussed in the literature\cite{vanLoock}, the spatial aspects of the problem 
has been ignored so far. In this paper we propose a full spatio-temporal
teleportation protocol. Our scheme allows us to teleport the
quantum state of the distributed in space-time electromagnetic field, which
can carry a spatial information like an optical image, or spatio-temporal
information like an animation or a movie.

\begin{figure}[h]
\leavevmode
\centering
\epsfxsize=8.3cm
\epsffile{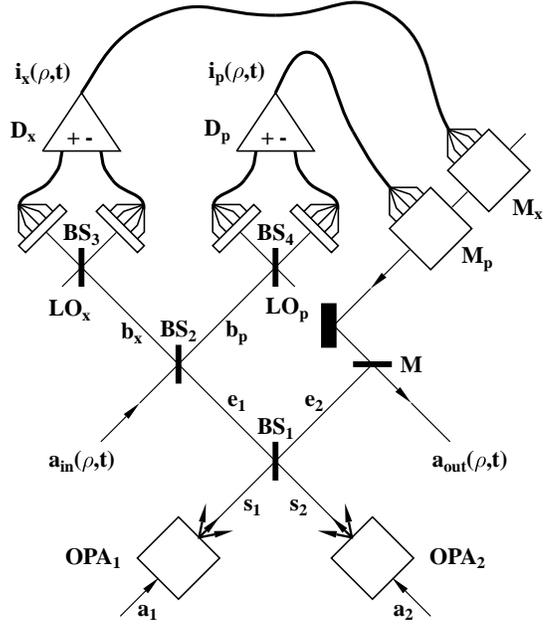}
\caption{Schematic of holographic teleportation...}
\label{fig1}
\end{figure}

Our teleportation scheme is similar to that described in \cite{Braunstein98a}
and
is shown in the figure.  The input light field to be teleported
from Alice to Bob is denoted by $A_{\rm in}(\r,t)$, where
$\r = (x,y)$ is the transverse coordinate. Two quadrature
components of the light field are detected ``point-by-point'' 
by two balanced homodyne detectors $D_x$ and $D_p$ formed by high efficiency 
multipixel photodetection matrices (CCD). The spatio-temporal quantum 
fluctuations of these quadrature components are locally 
imprinted into the photocurrents $i_x({\r},t)$ and $i_p({\r},t)$ on 
the output of individual pixels of CCD cameras. These photocurrents 
are sent from Alice to Bob via two multichannel parallel 
classical communication lines. Bob uses these photocurrents for reconstruction 
of the field $A_{\rm out}({\r},t) $ via two multichannel modulators 
$M_x$ and $M_p$ which modulate in space and time an incoming plane coherent 
light wave.

The essential part of the teleportation scheme is a pair of traveling-wave
optical parametric amplifiers OPA$_1$ and OPA$_2$ used for creation of two
broadband multimode entangled Einstein-Podolsky-Rosen (EPR) light beams. 
Owing to the multimode nature
of entanglement created by the OPAs, our scheme allows for parallel
teleportation with optimum fidelity of $N$ elements of the input wavefront,
preserving the space-time correlations between these elements. This number
is given by the ratio of the wavefront cross-section to the coherence area
of the light created by OPAs. In the generic teleportation scheme
\cite{Braunstein98a} $N=1$.

The EPR beams $E_n(\r,t)$, n = 1,2, are created by the
interference mixing at the 50:50 beam splitter BS$_1$ of the fields
$S_m(\r,t)$, m = 1,2, in broadband multimode squeezed state,
\BE
      E_n(\r,t) = \sum_{m=1,2} R_{nm} S_m(\r,t),
                     \label{mixing}
\EE
where
\BE
      \left\{ R_{nm} \right\} = \frac{1}{\sqrt{2}} \left(
      \begin{array}{rr}
      1 & 1 \\ -1 & 1 \\
      \end{array}
      \right),
                     \label{beamsplitter}
\EE
is the scattering matrix of the beam splitter BS$_1$. The light waves
$S_m(\r,t)$ in the broadband multimode squeezed state are created by two 
traveling-wave optical parametric amplifiers OPA$_1$ and OPA$_2$. The 
detailed description of properties of such squeezed light can be found, 
for example in Ref.~\cite{Kolobov99}. The transformation of the input 
fields $A_m(\r,t)$ in the
vacuum state into the output fields $S_m(\r,t)$ in the broadband multimode
squeezed state is described in terms of the Fourier components of these
operators in frequency and spatial-frequency domain,
\BE
     s_m(\vec{q},\Omega)=\int d\r \,dt \exp[i(\Omega t -\vec{q}\cdot \r\,)]
     S_m(\r,t),
               \label{Fourier}
\EE
and similar for $a_m(\vec{q},\Omega)$. These Fourier components are related as
follows, 	       
\BE
      s_m(\vec{q},\Omega) = U_m(\vec{q},\Omega) a_m(\vec{q},\Omega) +
      V_m(\vec{q},\Omega) a_m^{\dag}(-\vec{q},-\Omega),
                      \label{squeezing}
\EE
where the coefficients $U_m(\vec{q},\Omega)$ and $V_m(\vec{q},\Omega)$ depend on
the pump-field amplitudes of the OPAs, their nonlinear susceptibilities and the
phase-matching conditions~\cite{Kolobov99}.

The spatial and temporal scales of our teleportation scheme are determined
by the orientation angle $\psi_m(\vec{q},\Omega)$ of the major axis of the  
squeezing ellipse,
\BE
     \psi_m(\vec{q},\Omega) = \frac{1}{2}\arg\left\{U_m(\vec{q},\Omega)
     V_m(-\vec{q},-\Omega)\right\},
                       \label{psi}
\EE
and by the degree of squeezing $r_m(\vec{q},\Omega)$,
\BE
       e^{\pm r_m(\vec{q},\Omega)} = |U_m(\vec{q},\Omega)| \pm
       |V_m(\vec{q},\Omega)|.
                         \label{exponential}
\EE
In analogy to the single-mode EPR beams, the multimode
EPR beams are created if squeezing in both channels is
effective, and the squeezing ellipses are oriented in
orthogonal directions. For simplicity we shall assume that 
OPA$_1$ and OPA$_2$ have such properties that,
$$
      U_1(\vec{q},\Omega) = U_2(\vec{q},\Omega) \equiv U(\vec{q},\Omega),
$$
\BE
      V_1(\vec{q},\Omega) = - V_2(\vec{q},\Omega) \equiv
      V(\vec{q},\Omega).
                      \label{orthogonal}
\EE
This situation is realized, for example, for broadband multimode squeezing
produced by an OPA with type-II phase matching\cite{Kolobov91a}. In the latter
case two independent squeezed beams correspond to two orthogonal polarization
components of the field with the following properties, 
$$
      r_1(\vec{q},\Omega) = r_2(\vec{q},\Omega) \equiv r(\vec{q},\Omega),
$$
\BE
       \psi_1(\vec{q},\Omega) = \psi_2(\vec{q},\Omega) \pm \pi/2 \equiv
       \psi(\vec{q},\Omega).
                      \label{orthogonal_2}
\EE
Under these conditions the EPR fields are entangled for the frequencies 
$\Omega$ and spatial frequencies $\vec{q}$ within the phase matching
of the OPA.

For observation of two quadrature components of the input
field $A_{\rm in}(\r,t)$ the input beam is mixed with one of two EPR beams,
$E_1(\r,t)$, at the beam splitter BS$_2$ with the same scattering matrix
as in Eq.~(\ref{beamsplitter}). This gives the input fields of the balanced
homodyne detectors $D_x$, $D_p$ as
$$
    B_x(\r,t) = \frac{1}{\sqrt{2}}\big(A_{\rm in}(\r,t) + E_1(\r,t)\big),
$$
\BE
    B_p(\r,t) = \frac{1}{\sqrt{2}}\big(-A_{\rm in}(\r,t) + E_1(\r,t)\big).
                 \label{mixing_2}
\EE
These fields in turn are mixed with the local oscillator fields LO$_x$ and
LO$_p$ having complex amplitudes $B_0$ and $-i B_0$, where $B_0$ is 
real, at beam splitters BS$_3$ and BS$_4$ with the same 
scattering matrices as for BS$_1$ and BS$_2$. We shall
assume that pixels of the CCD matrices have the area much smaller than
the coherence area $S_{c}$ of the EPR beams. In this case it can be shown that 
the photocurrents collected from individual pixels of D$_x$ and $D_p$ at the 
point $\vec{\rho}$ are given by, 
$$
    i_x(\r,t) = B_0 \big(B_x(\r,t) + B^{\dag}_x(\r,t)\big),
$$
\BE
    i_p(\r,t) = B_0 \frac{1}{i} \big(B_p(\r,t) - B^{\dag}_p(\r,t)\big).
                   \label{current}
\EE
These photocurrents are sent from Alice to Bob via two
multichannel classical communication lines and are used by Bob for
local modulation of an external coherent wave,
phase matched with the squeezed fields \cite{Furusawa98,Polkinghorne99}.
In the modulated beam the field component
$\sim i_x(\r,t) -i i_p(\r,t)$ is created. The teleported field
$A_{\rm out}(\r,t)$ is obtained by interference mixing on the
mirror M with hight reflectivity of the modulated
field with the second EPR beam $E_2(\r,t)$,
\BE
       A_{\rm out}(\r,t) = E_2(\r,t) + g\big(i_x(\r,t) -i i_p(\r,t)\big).
                       \label{out_coord}
\EE
Here $g$ is the coupling constant which takes into account the efficiency of
modulation and the transmission of the mirror M.
The teleportation takes place when $ g B_0 \sqrt{2} = 1$. We
find the teleported field $A_{\rm out}(\r,t)$ in the form,
\BE
       A_{\rm out}(\r,t) = A_{\rm in}(\r,t)+F(\r,t),
                       \label{out_fourier}
\EE
where
\BE
      F(\r,t) = E_2(\r,t) + E_1^{\dag}(\r,t),
                       \label{noise}
\EE
is the noise field, added by the teleportation process.
In the ideal case of perfect entanglement of two EPR beams at all
frequencies $\Omega$ and spatial frequencies $\vec{q}$ the 
terms $E_2(\r,t)$ and $E_1^{\dag}(\r,t)$ are perfectly anticorrelated and
their quantum fluctuation cancel each other. This would correspond to the
perfect ``point-to-point'' in space and instantaneous in time teleportation of
the quantum state of the input field with an arbitrary distribution in space and 
time, $A_{\rm out}(\r,t) = A_{\rm in}(\r,t)$. However such
teleportation would require infinitely large energy of EPR beams. 
Indeed, firstly as in the single-mode case, one would have to achieve an 
infinite squeezing per single coherence volume of an EPR beam. Additionally,
since now we have broadband multimode entanglement, one would need an infinite
number of elementary coherence volumes in the EPR beams. In practice
teleportation will never be point-to-point in space and instantaneous in time
but always ``on average'' within some spatial area and within some finite time
interval.

In order to verify that teleportation has actually taken place and to evaluate
its quality we shall follow the strategy employed in Ref.~\cite{Furusawa98} and
introduce the third party, Victor. He will make two measurements with the
fields $A_{\rm in}(\r,t)$ and $A_{\rm out}(\r,t)$ and according to the results
of these measurements he will decide whether teleportation was successful or
not.

We shall assume that Victor performs his own homodyne detection measurement of
an arbitrary quadrature component of the {\sl in} and {\sl out} fields
determined by the angle $\phi$,
\BE
     i_{\rm out}^{(\phi)}(\r,t) = A_0 \big(A_{\rm out}(\r,t)e^{-i\phi}+
     A_{\rm out}^{\dag}(\r,t)e^{i\phi}\big),
                      \label{current_2}
\EE
where $A_0$ is the real amplitude of Victor's local oscillator. For evaluation
of the teleportation quality Victor observes the photocurrent noise spectrum of
the teleported field defined as,
$$
     \left(\delta i^2_{\rm out}\right)^{(\phi)}_{\vec{q},\Omega} = 
     \int d\r \,dt \exp[i(\Omega t -\vec{q}\cdot \r\,)]
$$     
\BE     
     \times
     \langle \delta i^{(\phi)}_{\rm out}(\vec{0},0) 
     \delta i^{(\phi)}_{\rm out}(\r,t)\rangle,
                      \label{spectrum_def}
\EE
where $\delta i^{(\phi)}_{\rm out}(\r,t)$ is the fluctuation of the 
photocurrent around its mean value. He compares this spectrum with an
analogous noise spectrum $\delta i^{(\phi)}_{\rm in}(\r,t)$ for the 
input field.
Using Eq.~(\ref{out_fourier}) and relations between the EPR fields and
the input fields of the OPAs we obtain,
$$
      \left(\delta i^2_{\rm out}\right)^{(\phi)}_{\vec{q},\Omega} =
      \left(\delta i^2_{\rm in}\right)^{(\phi)}_{\vec{q},\Omega} +
      2A_0^2\left\{e^{-2r(\vec{q},\Omega)}
      \cos^2\psi(\vec{q},\Omega) \right.
$$
\BE      
      \left.
      + e^{2r(\vec{q},\Omega)}
      \sin^2\psi(\vec{q},\Omega) \right\}.
                       \label{spectrum_res}
\EE
Let us consider first teleportation of a classical plane monochromatic wave,
i.~e.~when the Fourier component of $A_{\rm in}(\r,t)$ with $\vec{q} = 0$ and 
$\Omega = 0$ is in coherent state and all other components in vacuum state. 
In this case $\left(\delta i^2_{\rm in}\right)^{(\phi)}_{\vec{q},\Omega} 
= a_0^2$. Without squeezing, $r(\vec{q},\Omega)=0$, we obtain 
$\left(\delta i^2_{\rm out}\right)^{(\phi)}_{\vec{q},\Omega} =
3\left(\delta i^2_{\rm in}\right)^{(\phi)}_{\vec{q},\Omega}$. In this 
classical limit of teleportation \cite{Furusawa98} the three-dimensional 
noise spectrum of photocurrent is multiplied by the same factor 3, as 
the one-dimensional $\Omega$-dependent spectrum in the single-mode case.

Quantum teleportation superseding this classical limit is possible when (i)
there is an effective squeezing, $r(\vec{q},\Omega) \gg 1$, in a
certain region of frequencies $\vec{q}$ and $\Omega$, and (ii)
the orientation angle $\psi(\vec{q},\Omega)$ within this region
is $\psi(\vec{q},\Omega) \simeq 0$.

It was shown earlier that the $\Omega$-dependence of the angle 
$\psi(\vec{q},\Omega)$ can be compensated by the frequency-dependent 
refraction index of nonlinear medium~\cite{Crough88},
and the $\vec{q}$-dependence is to large extent compensated
by a thin lens properly inserted after the OPA~\cite{Kolobov89,Kolobov99}. 
Assuming that OPAs have phase matching degenerate in frequency and angle,
we can introduce the coherence area $S_{c} = (2\pi/q_{c})^2$ and the coherence 
time $T_{c} = 2\pi/\Omega_{c}$ for the EPR fields. The squeezing is 
effective for $|q| \leq q_{c}/2$ and $|\Omega| \leq \Omega_{c}/2$
The three-dimensional photocurrent noise spectra of the {\sl in} and {\sl out} 
fields become identical within this range of frequencies and spatial 
frequencies, $\left(\delta i^2_{\rm out}\right)^{(\phi)}_{\vec{q},\Omega} =
\left(\delta i^2_{\rm in}\right)^{(\phi)}_{\vec{q},\Omega}$. 
This corresponds to quantum teleportation which cannot be achieved without
broadband multimode EPR beams shared by Alice and Bob.

The low-frequency noise
suppression in Eq.~(\ref{spectrum_res}) means that if the photocurrents
are collected from the area $S \gg S_{c}$ during the time $T \gg
T_{c}$, the fluctuations in the {\sl in} and {\sl out}
measurements have similar statistics.
In the opposite case of measurement with $S \ll S_{c}$, $T \ll
T_{c}$ the high-frequency noise contribution in Eq.~
(\ref{spectrum_res}) degrades the quality of teleportation. Therefore, 
for the space-time distributed fields the concept of
teleportation fidelity must include coarse-grained description
with the scales $S_{c}$, $T_{c}$ of the EPR beams.

The teleportation fidelity is degraded due to the noise field
$F(\r,t)$ in Eq.~(\ref{out_fourier}). The commutation relations for
$F(\r,t)$ correspond to classical noise,
\BE
     [F(\r,t),F^{\dag}(\r\,',t')] = 0,
     \quad [F(\r,t),F(\r\,',t')] = 0.
            \label{commutation_noise}
\EE
The statistics of this noise field are determined in the most general form by
the characteristic functional,
$$
     {\cal F}(\lambda,\lambda^{\star}) = \langle \exp \Bigl\{\int d\r\, dt
     \Bigl(
     \lambda(\r,t) F^{\star}(\r,t)
     \Bigr. \Bigr.
$$      
\BE   
     \Bigl. \Bigl.
     - \lambda^{\star}(\r,t) F(\r,t)\Bigr) \Bigr\}
     \rangle.
                \label{functional_def}
\EE
Our calculations which will be published elsewhere give,
$$
     {\cal F}(\lambda,\lambda^{\star}) = \exp \Bigl\{-\int d\r\, d\r\,' dt\, dt'
     \lambda(\r,t) \lambda^{\star}(\r\,',t')
     \Bigr.
$$     
\BE   
      \Bigl.  
      \times G(\r-\r\,',t-t') \Bigr\},
                 \label{functional_res}
\EE
where the Fourier transform of the Green function $G(\r,t)$ reads, 
$$
     G(\vec{q},\Omega) = \left| U(\vec{q},\Omega) -
     V^{\star}(-\vec{q},-\Omega) \right|^2 
$$      
\BE     
     = e^{-2r(\vec{q},\Omega)}
     \cos^2\psi(\vec{q},\Omega) + e^{2r(\vec{q},\Omega)}
     \sin^2\psi(\vec{q},\Omega).
                  \label{green}
\EE
It follows from Eq.~(\ref{functional_res}) that the noise is
Gaussian and the correlation functions of the fields $F(\r,t)$ and 
$F^{\star}(\r,t)$ of an arbitrary order
are expressed in standard way via the second-order correlation functions
\BE
      \langle F(\r,t) F^{\star}(\r\,',t')\rangle = G(\r-\r\,',t-t').
                  \label{pair}
\EE
Incidentally, similar Green function describes the photocurrent correlations
in space-time by the homodyne detection of multimode
squeezed light~\cite{Kolobov91b}.

When squeezing and entanglement are not present, the Green function is 
$\delta$-correlated in space-time,
\BE
        G(\r,t) = \delta(\r\,) \delta(t).
                   \label{delta}
\EE
In presence of an effective entanglement with the scales $S_{c}$, $T_{c}$ 
the positive $\delta$-correlated term in Eq.~(\ref{delta}) is 
accompanied by a negative term due to spatio-temporal 
anticorrelations on the scales $S_{c}$, $T_{c}$. 
Consider the field variables, averaged over the pixel $S_j$ of area $S$ 
within the time window $T_i$ of duration $T$:
\BE
        F(j,i) = \frac{1}{\sqrt{ST}} \int_{S_j} d\r \int_{T_i} dt
                 F(\r,t).
                   \label{averaged}
\EE
The Green function for these averaged variables goes over to the covariance
matrix $\langle F(j,i) F^\star (j',i') \rangle$. For effective squeezing, 
$r(\vec{q},\Omega) \gg 1$, and large sampling volume, $S \gg S_c$, 
$T \gg T_c$, we obtain
\BE
        \langle F(j,i) F^\star (j',i') \rangle \rightarrow 
        \delta_{jj'} \delta_{ii'} e^{-2r(\vec{0},0)}.
                   \label{matrix}
\EE
These properties of the noise correlation function mean that
noise on the scales $S \ll S_{c}$, $T \ll T_{c}$ is not
eliminated, but on the scales $S \geq S_{c}$, $T \geq T_{c}$
effective entanglement results in significant noise suppression.

To conclude, we have proposed the protocol for quantum
teleportation of the distributed in space-time light fields. The
protocol is based on the entanglement between the corresponding
coherence volumes $cT_{c} S_{c}$ of the broadband multimode
EPR fields produced by two traveling-wave OPAs. 
Every such volume determines an elementary degree of
freedom of the input field, whose quantum state can be effectively
teleported, i.~e.~the ``resolving power'' of teleportation.

It follows from Eq.~(\ref{out_fourier}), which relates in
the Heisenberg picture the {\sl out} field to {\sl in} field, that
any average product of the {\sl out} fields, taken at arbitrary
space-time points, is equal to the analogous average product of
the {\sl in} fields plus average noise term which can be made small using
entangled EPR beams. That is, our protocol
preserves the space-time quantum correlations in the {\sl in} field.
Teleportation with these features can be called {\it quantum holographic 
teleportation}.

In fact, in our three-dimensional generalization of the continuous variable 
teleportation protocol \cite{Braunstein98a,Furusawa98} one can easily recognize
an extension to the quantum domain of the conventional non-stationary
holography. As in holography, the distributed in space-time input
field is mixed with the local oscillator waves. Classical
photocurrent densities $i_x(\r,t)$ and $i_p(\r,t)$ are equivalent to
non-stationary holograms, each for one of two quadrature
components. These holograms are transmitted via classical
multichannel communication lines and used for reconstruction
of both quadrature components of the input field by means of
non-stationary modulation of the beam from an external laser. The
novel feature, which converts the holography to the quantum
holographic teleportation is a pair of broadband multimode EPR beams shared by
Alice and Bob.

\section*{Acknowledgments}

This work was supported by the Network QSTRUCT of the TMR program of the
European Union and by the Russian Foundation for Basic Research
Project 98-02-18129.


\end{document}